# Theoretical investigation on the ferromagnetic two-dimensional scandium monochloride sheet that has a high Curie temperature and could be exfoliated from a known material

Guo Wang*, Xinyu Wang and Yi Liao

*Department of Chemistry, Capital Normal University, Beijing 100048, China.*

*Email:* wangguo@mail.cnu.edu.cn

ABSTRACT: A two-dimensional scandium monochloride sheet was investigated by using density functional theory. It could be exfoliated from a known bulk material with a cleavage energy slightly lower than that of graphene. The sheet has a ferromagnetic ground state with a Curie temperature of 100 K. Moreover, the sheet becomes a half-metal under hole doping. The Curie temperature increases to 250 K with the doping amount of 0.4 per primitive cell, which is close to the ice point. The two-dimensional scandium monochloride sheet should be a good candidate for two-dimensional spintronics.

## 1. Introduction

Spintronic devices, in which spins play the role of charge carriers, are promising candidates for next generation information technology. Compared with traditional electronic devices, spintronic devices take the advantage of low power consumption [1] and should have wide applications in integrated circuits. Meanwhile, the successful exfoliation of graphene [2] opens an avenue for two-dimensional materials. Two-dimensionality should further reduce the device size to accommodate the Moore's law. For a typical spintronic device, a magnetic semiconductor in the center and two magnetic metal (preferential half-metal) electrodes at the two ends are required. However, ferromagnetic (FM) materials are rare compared to nonmagnetic or antiferromagnetic (AFM) ones. Two-dimensionality also provides a way for searching FM materials [3-9] with high Curie temperature. For easily exfoliated van der Waals materials, weak interlayer magnetic coupling results in low Curie temperature even if the material is FM. When exfoliated from bulk materials, stronger intralayer couplings become dominate and Curie temperature can increase. During recent years, two-dimensional spintronics have attracted much attention. In the last year, $CrI_3$ monolayer with long-range FM ordering was successfully fabricated [10]. Although the Curie temperature is only 45 K, it demonstrates that the two-dimensional ferromagnetism is feasible. In fact, the first FM material discovered in 1960 is also a chromium halide $CrBr_3$, with Curie temperature of only 37 K [11]. Weak interlayer halogen interaction in layered halides should make the exfoliation feasible. Actually, two-dimensional $CrX_3$ (X=Cl, Br, I) [12,13] has been predicted before the successful fabrication. In addition, several FM first-row transition metal halides were predicted [12-20]. The transition metal atoms have low valence, because there is no 3d electron left in a metal atom with the highest valence. Among these metal halides, some of the structures such as $CrX_3$ and $VCl_3$ have low cleavage energies compared to that of graphene [12,15], while some of the structures such as CrI$_3$, CrCl$_3$, VI$_3$ and VCl$_3$ have high Curie temperatures of 95 (107), 49 (66), 98 and 80 K, respectively [12,13,16]. Usually, chlorides should be more stable than the iodides. For example, the operation of CrI$_3$ requires a glove box [10]. In the present work, a FM two-dimensional early 3d transition metal chloride with low valence, scandium monochloride (ScCl), was investigated theoretically. The calculations indicated that the two-dimensional ScCl sheet has a low cleavage energy compared to that of graphene and has a high Curie temperature of 100 K. Furthermore, the Curie temperature can be increased to 250 K for a stable hole doped structure.

## 2. Models and computational details

As shown in Fig. 1(a), two Sc layers are sandwiched between two Cl layers in the two-dimensional ScCl sheet. Each primitive cell contains two Sc and two Cl atoms. Besides the FM configuration, three AFM configurations called AFM1, AFM2 and AFM3 shown in Fig. 1(b)-1(d) were proposed. A hexagonal primitive cell is enough for calculating the FM or AFM1 configuration while a 2×1 supercell is needed for the AFM2 and AFM3 configurations. The Perdew-Burke-Ernzerhof (PBE) [21] density functional is used for the geometrical optimization and electronic properties calculations. The projector-augmented wave basis in the Vienna ab initio simulation package was used. The GGA+U method with various U values for 3d electrons of Sc atoms was applied for test purpose. The HSE06 hybrid density functional [22], which can accurately describe the band gaps of solids [23], was adopted to confirm the results. Unless explicitly stated, the results below are all calculated by the PBE density functional. Monkhorst-Pack sampling with 41×41×1 k-points was used for the primitive cell, so that the mesh density is about 0.05 Å$^{-1}$. Further increasing the mesh density has little effect on the calculated results. A vacuum layer with thickness of about 15 Å was used along the z direction (vertical to the two-dimensional plane) to avoid interactions between image structures. The phonon dispersions were calculated based on the density functional perturbation theory with the aid of Phonopy code [24]. Ab initio molecular dynamics simulations were performed. A canonical ensemble was simulated using the algorithm of Nosé. The SMASS parameter was set to 4 so that the estimated Nosé frequency is close to the phonon frequencies.

## 3. Results and discussions

*3.1. Structures and electronic properties*

The optimized lattice constant of the two-dimensional ScCl sheet with the FM configuration is 3.48 Å. The total magnetic moment is 1.52 $\mu_B$. Detailed analysis indicates that the magnetic moment is mainly contributed by the d orbitals of the Sc atoms. The spin density shown in Fig. 2 clearly illustrates this point. To confirm the ground state, structures with AFM configurations were also geometrically optimized. The relative energies of the FM, AFM1, AFM2 and AFM3 configurations denoted in Fig. 1(b)-1(d) are 0, 42, 49 and 38 meV per primitive cell. This indicates that the two-dimensional ScCl sheet has a FM ground state. The band structures of the FM configuration shown in Fig. 3(a) indicate that the two-dimensional ScCl sheet is a FM metal. There are bands crossing the Fermi level for the both spins and the band structures are not degenerate. Moreover, a gap exists about 0.1 eV below the Fermi level for the minority spin. To confirm the band structures calculated by the PBE density functional, the band structures were also calculated by the HSE06 hybrid density functional as shown in Fig. 3(b). The band shapes calculated by the two methods are quite similar. The main difference is that the energy of each band calculated by the HSE06 hybrid density functional is lower than that calculated by the PBE density functional. This does not affect the conclusion that the two-dimensional ScCl sheet is a FM metal. Also, there is a gap below the Fermi level in the band structures calculated by the HSE06 hybrid density functional. Moreover, the total magnetic moment is 1.61 $\mu_B$, close to the value calculated by the PBE density functional. This confirms the results calculated by the PBE density functional. Furthermore, GGA+U calculations with U values from 1 to 5 eV were performed for test purpose. After the calculations, the band shapes do not change much. The gap below the Fermi level for the minority spin is taken as a descriptor. It is 0.58 or 0.43 eV calculated by the HSE06 or PBE density functional. The gap is 0.34, 0.20 or 0.07 eV when the U equals to 1, 2 or 3 eV. The gap disappears when the U equals to 4 and 5 eV. The total magnetic moments are 1.49, 1.44, 1.40, 1.36 and 1.33 $\mu_B$ for the five U values. Taken the results calculated by the HSE06 density functional as benchmarks, the gap and total magnetic moment calculated by the PBE density functional are the closest. Therefore, the GGA+U method will not be used below.

It is noted above that there is a gap about 0.1 eV below the Fermi level for the minority spin as shown in Fig. 3(a). Hole doping should be an effective method for suppressing the states at the Fermi level for the minority spin. Following this idea, hole doped two-dimensional ScCl sheets were investigated. Hole doping amounts from 0.1 to 1 per primitive cell were considered. With the hole doping amount of 0.1, the states at the Fermi level for the minority spin has not been suppressed. From the amount of 0.2, the sheet becomes a half-metal. Compared to the FM configuration, the AFM1 configurations have the energies of 95, 104, 97, 78, 63, 50, 38, 27 and 19 meV per primitive cell when the hole doping amounts are 0.2, 0.3, 0.4, 0.5, 0.6, 0.7, 0.8, 0.9 and 1. The corresponding values are 100, 120, 124, 116, 106, 97, 86, 73 and 60 meV for the AFM2 configurations and 69, 85, 87, 82, 77, 77, 75, 66 and 53 meV for the AFM3 configurations. These indicate that the FM configurations are the ground states and the nine doped sheets are all half-metals. Since their band structures are similar, those of the two-dimensional ScCl sheet doped with 0.4 hole are taken as examples and are shown in Fig. 3(c). By rational modifying the band structures, the 100% spin polarization can be achieved. Among these sheets, the AFM configurations are the least stable with 0.3 or 0.4 hole doping. For the 0.3 hole doped sheet, the energies of the three AFM configurations are 104, 120 and 85 meV, respectively. These are 97, 124 and 87 meV for the 0.4 hole doped sheet. Considering the lattice constant, the two hole doping amounts correspond to hole doping density of $2.9 \times 10^{14}$ and $3.8 \times 10^{14}$ cm$^{-2}$. These are much smaller than the value $2 \times 10^{15}$ cm$^{-2}$ realized in experiment using electrolyte as a gate dielectric [25]. Thus the doping could be feasible.

*3.2. Curie temperature*

Since three AFM configurations were considered, three types of magnetic couplings between the Sc atoms were investigated. As shown in Fig. 1(b), three exchange parameters $J_1$, $J_2$ and $J_3$ denote the couplings between the nearest two Sc atoms in the same layer, the nearest and next-nearest two Sc atoms in the different layers. The three distances are 3.48, 3.24 and 4.76 Å in an undoped sheet. According to the Heisenberg model, the Hamiltonian is

$$H = -\sum_{i,j} J_1 S_i \cdot S_j - \sum_{k,l} J_2 S_k \cdot S_l - \sum_{m,n} J_3 S_m \cdot S_n$$

in which $S$ (about 0.4, considering the total magnetic moment) is the spin of a Sc atom. The energy functionals of the configurations are written as:

$$E_{FM} = (-6J_1 - 3J_2 - 3J_3)S \cdot S + E_0$$

$$E_{AFM1} = (-6J_1 + 3J_2 + 3J_3)S \cdot S + E_0$$

$$E_{AFM2} = (2J_1 + J_2 - 3J_3)S \cdot S + E_0$$

$$E_{AFM3} = (2J_1 - J_2 + 3J_3)S \cdot S + E_0$$

From the energies of the FM and AFM configurations, the exchange parameters can be calculated:

$$J_1 = (-E_{FM} - E_{AFM1} + E_{AFM2} + E_{AFM3}) / 16 S \cdot S$$

$$J_2 = (-E_{FM} + E_{AFM1} + E_{AFM2} - E_{AFM3}) / 8 S \cdot S$$

$$J_3 = (-E_{FM} + E_{AFM1} - 3E_{AFM2} + 3E_{AFM3}) / 24 S \cdot S$$

The three exchange parameters are 16, 47 and 4 meV for the undoped sheet. The $J_3$ is the smallest due to the long distance. To obtain the Curie temperature, it is generally accepted that the Monte Carlo simulations give better estimation than the mean-field approximation. An 80×80 supercell was constructed so that the number of metal atoms is in the order of $10^4$. The spin flips randomly during the $10^8$ simulation steps at different temperatures. As shown in Fig. 4, the total magnetic moment firstly decreases gently with the temperature. Then it deceases significantly near the 50% value. Finally it slowly decreases to near zero. The temperature when the total magnetic moment reduces to about 50% is taken as the Curie temperature. The undoped sheet has a Curie temperature of 100 K. This value is close to the values 95 (107) and 98 K for $CrI_3$ [12,13] and $VI_3$ [16] obtained in some previous studies, in which three exchange parameters were also considered in the Monte Carlo simulations. The Curie temperature 100 K of the two-dimensional ScCl sheet is larger than the values 49 (66) and 80 K for the stable [10] chlorides $CrCl_3$ [12,13] and $VCl_3$ [16].

Since the AFM configurations are less stable with 0.3 or 0.4 hole doping, the Curie temperatures for the two cases were also investigated. The three exchange parameters are 40, 107 and 1 meV for 0.3 hole doping while 45, 107 and −3 meV for 0.4 hole doping. The $J_1$ and $J_2$ are much larger than those of the undoped sheet and the $J_3$ is also quite small. Monte Carlo simulations were also performed for the doped sheets. As shown in Fig. 4, the Curie temperature increases to 230 or 250 K for the 0.3 or 0.4 hole doped two-dimensional ScCl sheets. The value 250 K is quite close to the ice point and is much higher than the temperature of dry ice. Thus it is favorable to real applications in two-dimensional spintronics. The increment of Curie temperature by doping can be explained by the density of states at the Fermi level. The value for the majority spin is 0.49, 0.64 or 1.13 eV$^{-1}$ per primitive cell for the undoped, 0.3 or 0.4 hole doped sheet. When excess carriers are doped into the structure, the density of states at the Fermi level increases. Electrons that make pairs bear stronger repulsion, so they tend to occupy different orbitals. The FM configurations become more stable. Then the larger exchange parameters result in higher Curie temperatures.

*3.3. Cleavage energy*

The two-dimensional ScCl sheet has a FM ground state with Curie temperature higher than other chlorides, such as CrCl$_3$ and VCl$_3$. Now it is time to examine whether it can be exfoliated from known bulk materials and whether the sheet is stable. The cleavage energy is an important parameter that affects the feasibility of exfoliation from bulk materials. Bulk ScCl has already been synthesized in 1977 [26]. In order to confirm the ground states of the bulk ScCl, an AFM4 configuration in which AFM couplings exist along the z direction between the Sc atoms was calculated in addition to the three AFM configurations described above. Since there is interlayer weak interaction in the bulk structure, a van der Waals density functional optPBE-vdW [27,28] was adopted to calculate the three-dimensional bulk structure and also the two-dimensional sheet for comparison. The calculations indicate that the AFM4 and FM configurations have similar energies. The energy of the FM configuration is about only 2 meV per two Sc and two Cl atoms lower. The FM and AFM4 configurations are nearly degenerate. Thus the Curie temperature of the bulk ScCl should be extremely low. The calculated cleavage energy is 0.30 J·m$^{-2}$. This is slightly lower than the value 0.37 J·m$^{-2}$ for graphene [29] and is similar to the values for CrCl$_3$ or VCl$_3$ calculated by a van der Waals density functional or the HSE06 density functional [12,15]. This implies that the two-dimensional ScCl sheet could be exfoliated from the known bulk ScCl.

*3.4. Phonon dispersion and ab initio molecular dynamics*

In order to check the dynamic stability, phonon dispersions were calculated for the two-dimensional undoped, 0.3 hole and 0.4 hole doped ScCl sheets. A 5×5×1 supercell was used for each structure. As shown in Fig. 5, no non-trivial imaginary frequency exists in the phonon dispersion. There are three small imaginary frequencies near the center of the Brillouin zone for all the three sheets. The largest imaginary frequency is only about 4 cm$^{-1}$. The small imaginary frequencies do not reflect the instability, but is affected by the numerical integration and the size of the supercell [30]. Detailed analysis indicates that the lowest three curves in each phonon dispersion correspond to three translation modes [31,32]. Thus the three sheets are dynamically stable. Ab initio molecular dynamics simulations were performed to verify the stability against heat. In Fig. 6, the simulation temperature is 300 K. The energy oscillates with the simulation time for several times, indicating that equilibrium was reached. The total magnetic moment remains unchanged during the simulation. After the simulation, the geometry shown in the inset of Fig. 6 has only a small distortion. These confirm the stability of the two-dimensional ScCl sheet.

## 4. Conclusions

A two-dimensional ScCl sheet was investigated by using density functional theory.

It has a FM ground state and the total magnetic moment is 1.52 $\mu_B$. Compared with the FM configuration, the AFM1, AFM2 and AFM3 configurations have the energies of 42, 49 and 38 meV per primitive cell. The exchange parameters for the nearest two Sc atoms in the same layer, the nearest and next-nearest two Sc atoms in the different layers are 16, 47 and 4 meV, respectively. The Curie temperature is 100 K obtained by Monte Carlo simulations. This is higher than those of $CrCl_3$ and $VCl_3$. In its band structures, there is a gap about 0.1 eV below the Fermi level for the minority spin. Thus hole doping can suppress the states at the Fermi level for the minority spin. The sheet becomes a half-metal at a wide doping range of 0.2-1 per primitive cell. The doping amount is far less than that already realized in experiments. The doping makes the FM configuration more stable and the Curie temperature increases to 230 or 250 K for the 0.3 or 0.4 hole doped two-dimensional ScCl sheets. The value 250 K is close to the ice point and higher than the temperature of dry ice. Phonon dispersions and ab initio molecular dynamics confirm its stability. Moreover, the sheet should be easily exfoliated from a known bulk ScCl with a cleavage energy slightly lower than that of graphene. Therefore, the two-dimensional ScCl sheet should be a good candidate for two-dimensional spintronics.

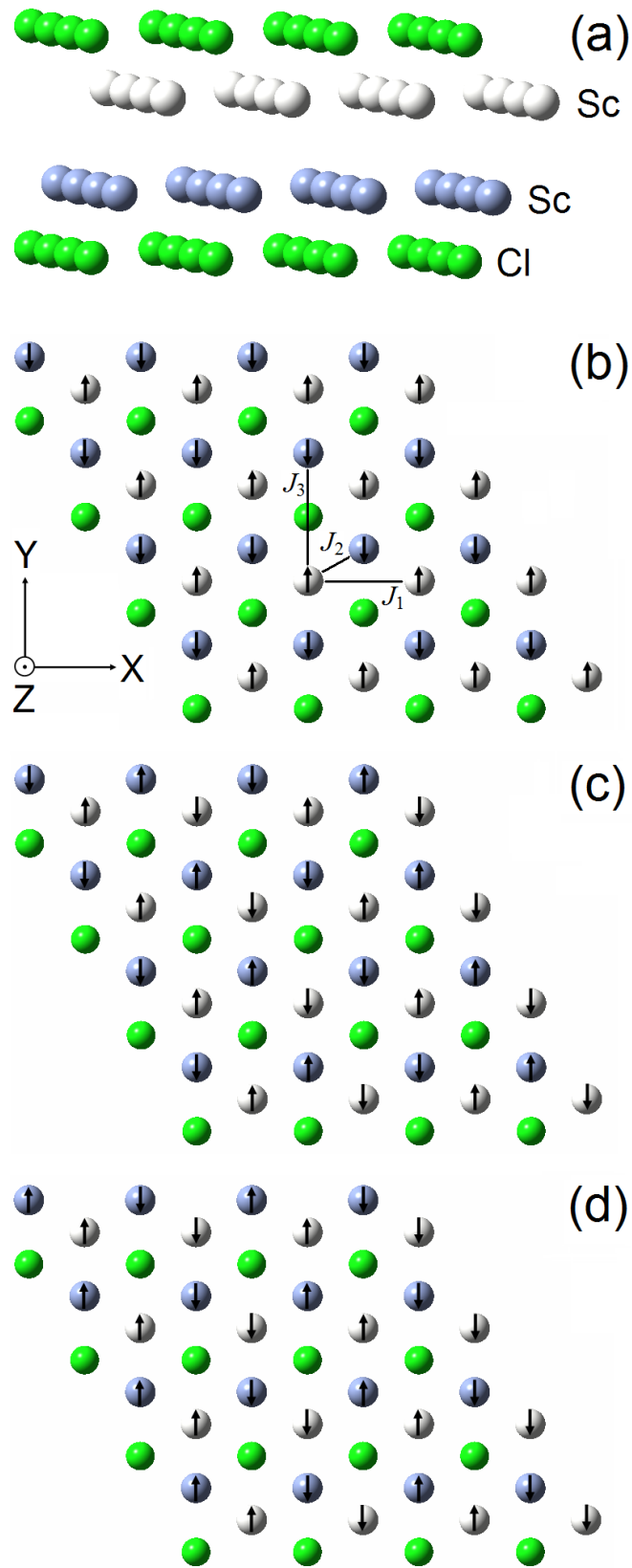

**Fig. 1** (a) Sideview and (b)-(d) top view of two-dimensional ScCl sheet, (b) AFM1, (c) AFM2 and (d) AFM3 configurations.

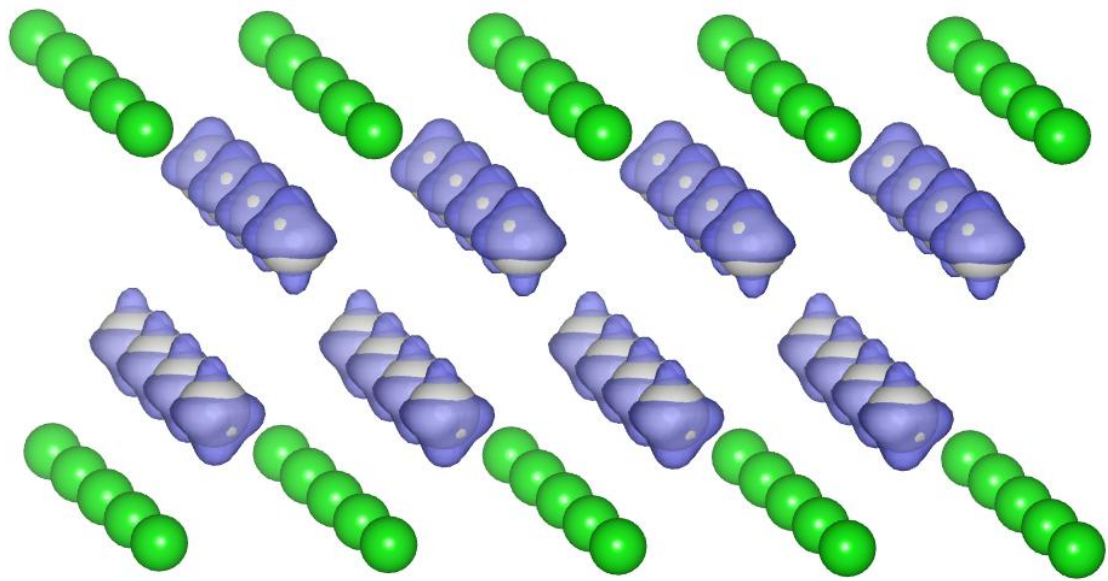

**Fig. 2** Spin density of two-dimensional ScCl sheet.

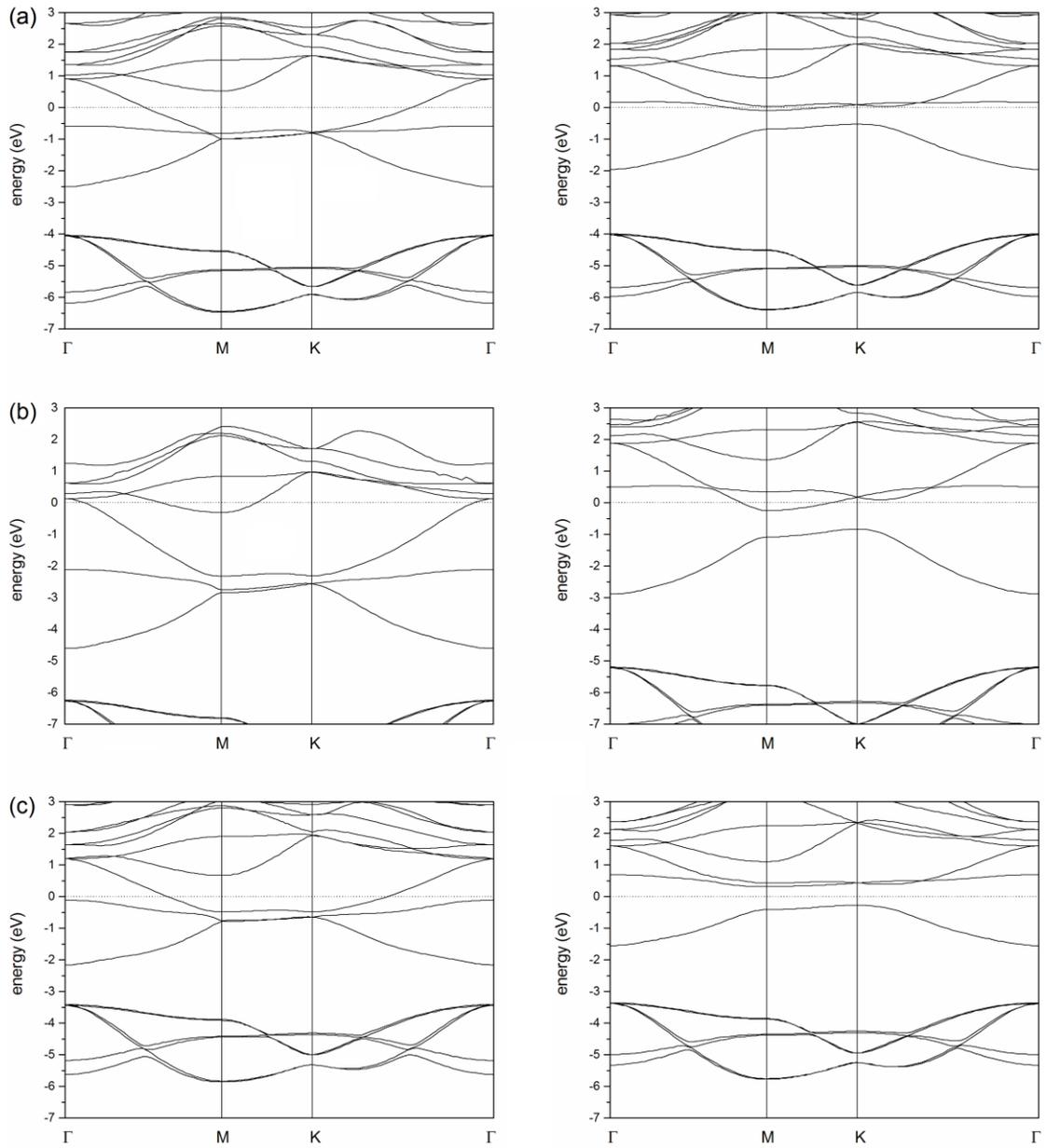

**Fig. 3** Band structures of FM configuration calculated with (a) the PBE and (b) the HSE06 density functionals, (c) band structures of FM configuration with 0.4 hole doping calculated with the PBE density functional. The left is for the majority spin and the right is for the minority spin. The Fermi level is set to zero.

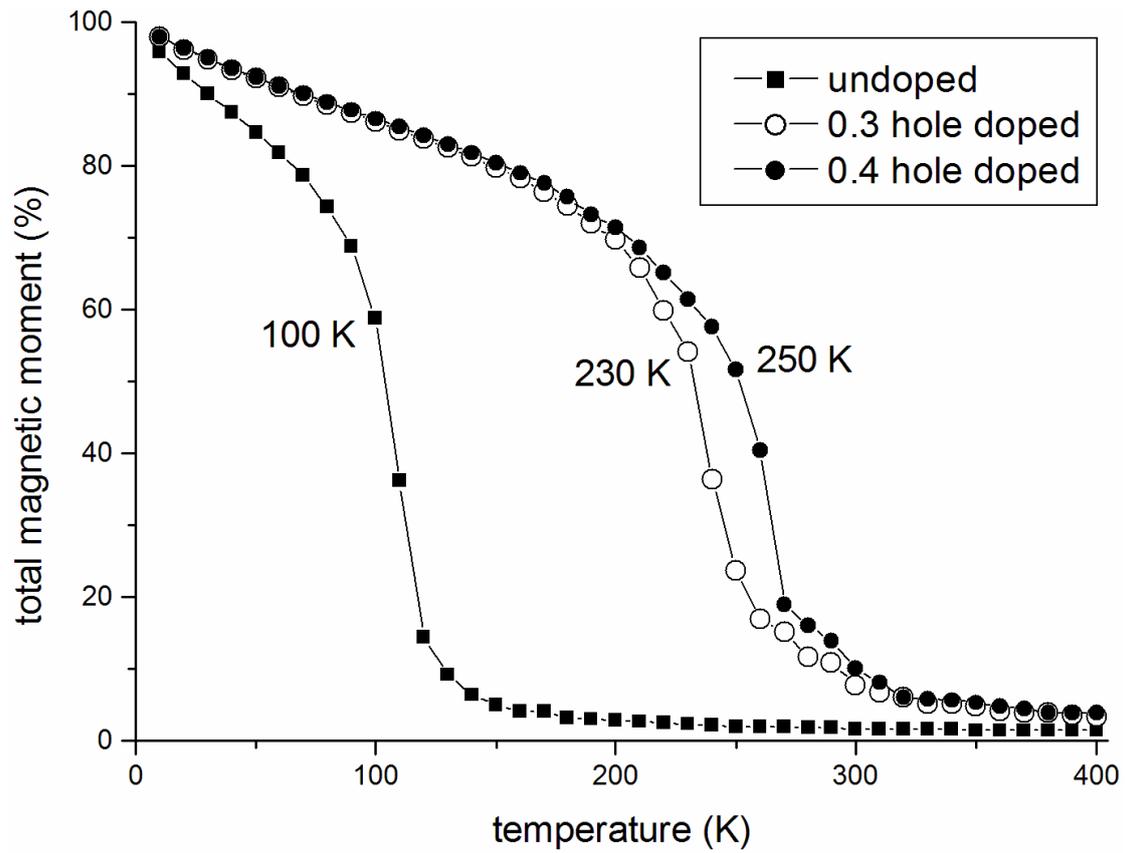

**Fig. 4** Curie temperatures of the undoped, 0.3 and 0.4 hole doped two-dimensional ScCl sheets obtained from Monte Carlo simulations.

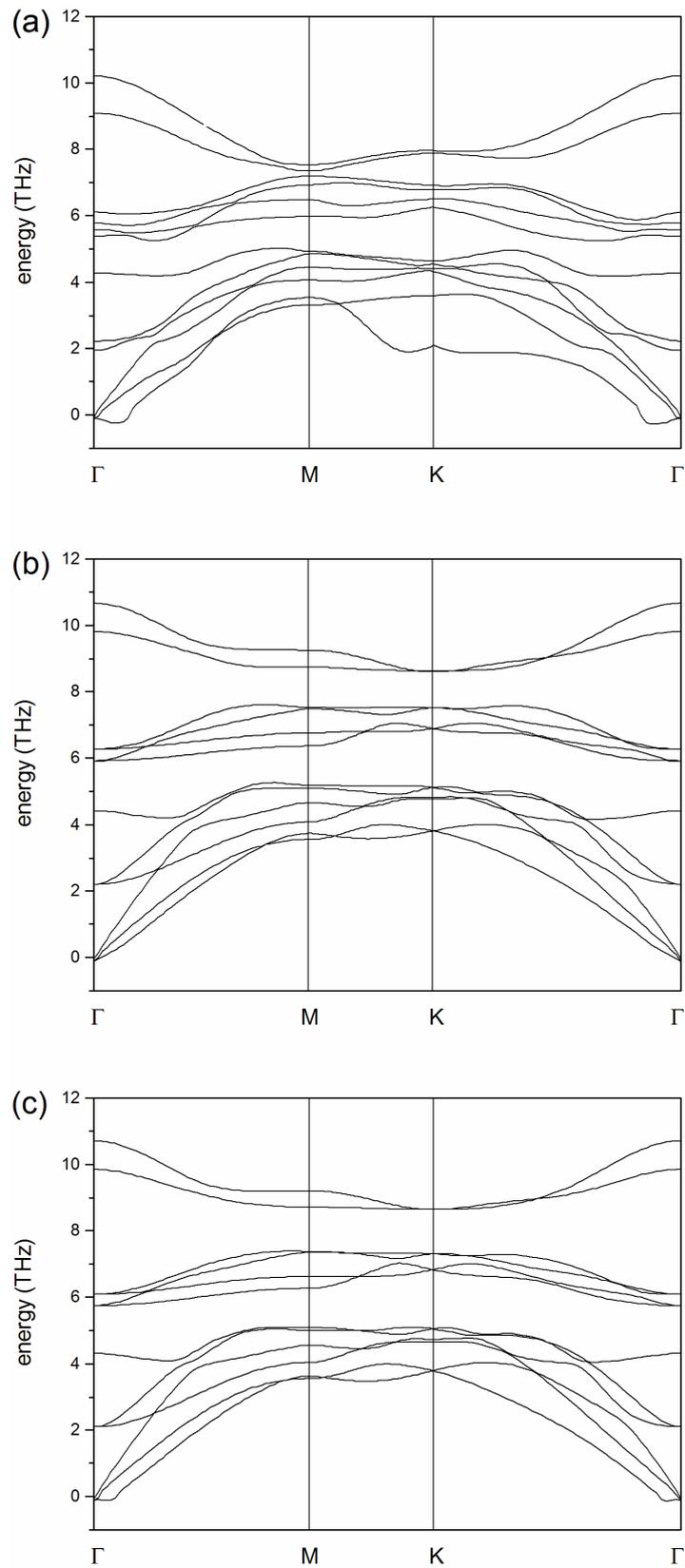

**Fig. 5** Phonon dispersions of (a) undoped, (b) 0.3 hole doped and (c) 0.4 hole doped two-dimensional ScCl sheets.

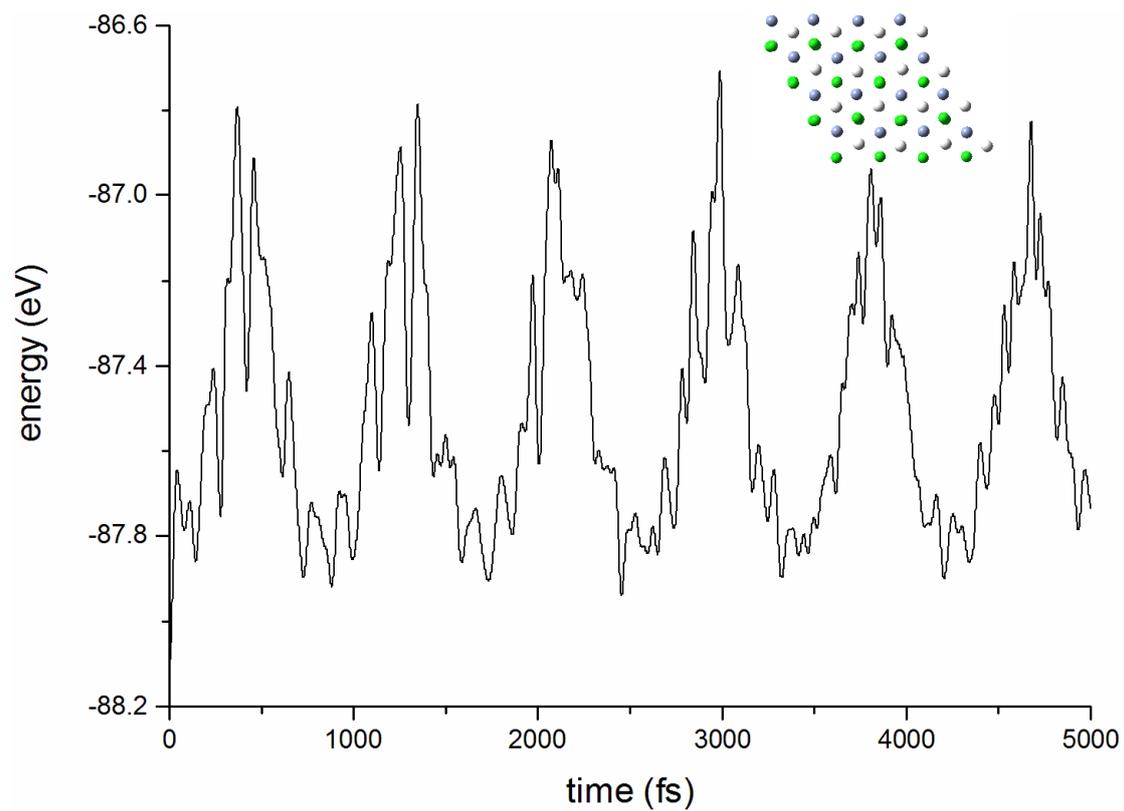

**Fig. 6** Ab initio molecular dynamics simulation of two-dimensional ScCl sheet. The inset is the geometry after 5000 steps simulation.